\documentclass[fleqn,usenatbib]{mnras}

\usepackage{newtxtext,newtxmath}
\usepackage{CJK}

\usepackage[T1]{fontenc}

\DeclareRobustCommand{\VAN}[3]{#2}
\let\VANthebibliography\thebibliography
\def\thebibliography{\DeclareRobustCommand{\VAN}[3]{##3}\VANthebibliography}

\usepackage{graphicx}	
\usepackage{amsmath}	

\title[The outbursts of GPSV16]{The long-term outburst(s) of GPSV16: from an intermediate to a FUor classification}

\author[C. Contreras Pe\~{n}a et al.]{
Carlos Contreras Pe\~{n}a,$^{1,2}$
Jeong-Eun Lee (이정은),$^{1,3}$
Philip W. Lucas,$^{4}$\thanks{E-mail: p.w.lucas@herts.ac.uk (PWL)}
Gregory Herczeg,$^{5,6}$
\newauthor
Doug Johnstone,$^{7,8}$
Zhen Guo,$^{9,10}$
Ho-Gyu Lee (이호규),$^{11}$
Hwan-Ki Kim (김환기),$^{1}$
\newauthor
Jessy Jose,$^{12}$
Mizna Ashraf$^{12}$
and Calum Morris$^{9}$
\\
$^{1}$Department of Physics and Astronomy, Seoul National University, 1 Gwanak-ro, Gwanak-gu, Seoul 08826, Korea\\
$^{2}$Research Institute of Basic Sciences, Seoul National University, Seoul 08826, Republic of Korea\\
$^{3}$SNU Astronomy Research Center, Seoul National University, 1 Gwanak-ro, Gwanak-gu, Seoul 08826, Korea\\
$^{4}$Centre for Astrophysics Research, University of Hertfordshire, College Lane, Hatfield, AL10 9AB, UK\\
$^{5}$Kavli Institute for Astronomy and Astrophysics, Peking University, Yiheyuan Lu 5, Haidian Qu, 100871 Beijing, Peoples Republic of China\\
$^{6}$Department of Astronomy, Peking University, Yiheyuan 5, Haidian Qu, 100871 Beijing, China\\
$^{7}$NRC Herzberg Astronomy and Astrophysics, 5071 West Saanich Rd, Victoria, BC, V9E 2E7, Canada\\
$^{8}$Department of Physics and Astronomy, University of Victoria, Victoria, BC, V8P 5C2, Canada\\
$^{9}$Instituto de F{\'i}sica y Astronom{\'i}a, Universidad de Valpara{\'i}so, ave. Gran Breta{\~n}a, 1111, Casilla 5030, Valpara{\'i}so, Chile\\
$^{10}$Chinese Academy of Sciences South America Center for Astronomy (CASSACA), National Astronomical Observatories, CAS, Beijing 100101, China\\
$^{11}$Korea Astronomy and Space Science Institute, 776 Daedeok-daero, Yuseong, Daejeon 34055, Korea\\
$^{12}$Indian Institute of Science Education and Research (IISER) Tirupati, Rami Reddy Nagar, Karakambadi Road, Mangalam (P.O.), Tirupati 517 507, India
}

\date{Accepted XXX. Received YYY; in original form ZZZ}

\pubyear{\the\year{}}

\begin{document}
\begin{CJK}{UTF8}{mj}
\label{firstpage}
\pagerange{\pageref{firstpage}--\pageref{lastpage}}
\maketitle

\begin{abstract}
FU Ori outbursts are thought to play a key role in stellar mass assembly and in the chemistry of protoplanetary disks during the early formation of stars. However, uncertainties remain regarding the universality of these events and the physical mechanism driving the high-amplitude variability. In this work, we present an analysis of optical, near- and mid-IR photometry (ZTF, UKIDSS GPS, NEOWISE) and near-IR spectra (IRTF, Gemini) of the eruptive variable Class I YSO GPSV16. The YSO, associated with the HII region G71.52$-$00.38 ($d=3.61$~kpc), showed two outbursts, one with $\Delta K_{\rm s}=2.2$~mag  (2005-2012) and a second starting in 2016 with $\Delta K_{\rm s}=5.6$~mag and accretion luminosity of $\sim$130 L$_{\odot}$. The outbursts displayed distinct spectroscopic characteristics: the first showed emission lines associated with a hot inner disk surface, whereas the second showed absorption lines arising from the cooler upper layers of a viscously heated disk. These features likely arose due to the different accretion rates reached during each outburst. The second outburst showed a two-stage mid-IR rise, requiring $\approx8.4$ years to reach peak brightness. The mid-IR rise also started 8 years before the onset of the optical outburst. The wavelength-dependent light curve points to an instability that is triggered at larger distances within the accretion disk and propagates inward. Assuming a propagation time of 8 years for the accretion wave, we estimate that the second outburst started at a distance of $r\sim0.4$~AU. These results show how long-term, multi-wavelength photometric monitoring can help identify the disk instabilities that trigger eruptions in YSOs.
\end{abstract}




\section{Introduction}\label{sec:intro}
\end{CJK}
FU Ori eruptions are accretion bursts up to $\sim10^{-4} $~M$_{\odot}$~yr$^{-1}$ that can last for over one hundred years \citep[see review by][]{1996Hartmann}. These 
outbursts (a.k.a FUors) are thought to play a critical role in the growth of stars and the chemistry of protoplanetary disks \citep[see review by][]{2023Fischer}. Historically very rare (about 10 known as of 2010), the past decade of monitoring surveys \citep[PTF/ZTF, {\it Gaia}, ASAS-SN, VVV/VVVX, and NEOWISE, ][]{2011Covey,2017Contreras_a,2018Hillenbrand,2023Nagy,2023Contreras_b,2024Guo} have revealed roughly 50 more FU Ori or FU Ori-like bursts \citep{2025Contreras_b}.   

There are still many unanswered questions regarding FUor outbursts, especially regarding the physical mechanism driving the enhanced accretion. A number of theoretical models have been proposed to explain the instabilities that lead to an outburst. The various invoked mechanisms include thermal viscous instability \citep{2024Nayakshin}, disk fragmentation \citep{2015Vorobyov}, gravitational and magnetorotational instabilities \citep{2009Zhu,2014Bae,2020Kadam}, planet-disk interaction \citep{2004Lodato}, extreme evaporation of planets \citep{2023Nayakshin}, binary interactions \citep{1992Bonnell}, cloudlet capture \citep{2019Dullemond}, and stellar flybys \citep{2000Reipurth,2022Dong}.

Long-term, multi-wavelength photometric monitoring can help identify the mechanism driving the outbursts. For example, magnetically driven accretion outbursts that are triggered at large distances within the accretion disk and then propagate inwards, will lead to mid- to near-IR rises years to decades before the onset of the optical brightening \citep{2023Cleaver, 2025Masley}. The detection of these infrared precursors, therefore, provides constraints on the locations where the outbursts initiate. Gaia17bpi and Gaia18dvy are two YSOs where a mid-IR outburst began $\sim$ 1-2 years before the onset of the optical outburst \citep{2018Hillenbrand,2020Szegedi-Elek}.

Here we report the discovery of a new FUor outburst where the mid-IR brightening began $\sim5-7$ years before the optical outburst. The YSO GPSV16 ($\alpha=$20:13:10.3, $\delta=+$33:31:28.2) is a known near-IR high-amplitude variable located 170\arcsec~from HII region G71.52-00.38  \citep{2016Xu, 2019Reid}. Based on its $2-22 \mu$m spectral index ($\alpha=0.6$), the source is classified as a Class I YSO in \citet{2017Lucas}. However, caution is needed as $\alpha$ is estimated from observations taken during outburst. The distance to the  HII region is estimated at $3.61^{+0.18}_{-0.16}$~kpc \citep{2016Xu}. The variability was first identified from comparing K-band photometry from 2005 and 2008, obtained by the UKIDSS Galactic Plane Survey \citep[GPS,][]{2008Lucas}. The source became $\sim2.5$ mag brighter in the 2008 epoch. Further photometric and spectroscopic monitoring confirmed that GPSV16 is an eruptive YSO with an intermediate, or PVM\footnote{We note that PVM is a new designation provided by \citet{2025Contreras_b} to these intermediate-type sources, and used in the classification catalogue of eruptive YSOS (OYCAT). The designation arises from a combination of names commonly given to these objects in the literature, Peculiar/V1647 Ori-like/MNors - PVM \citep[see][for further discussion on this topic]{2025Contreras_b}.}, classification \citep{2014Contreras}. 

In \S \ref{sec:obs} we describe the photometry and spectroscopy used in this paper.  In \S \ref{sec:sfr} we use near-IR $JHK$ observations to indicate the likely association of GPSV16 with HII region G71.52-00.38. In \S \ref{sec:outbursts} we describe the photometry and spectroscopy of the two outbursts observed in this source. In \S \ref{sec:secout} we focus on the properties of the second and largest outburst of GPSV16. These include, determination of its luminosity and the analysis of the two-stage brightening observed in the mid-IR. Finally, in \S \ref{sec:summary} we present a summary of our results.








\begin{figure*}
	\resizebox{\textwidth}{!}{\includegraphics[angle=0]{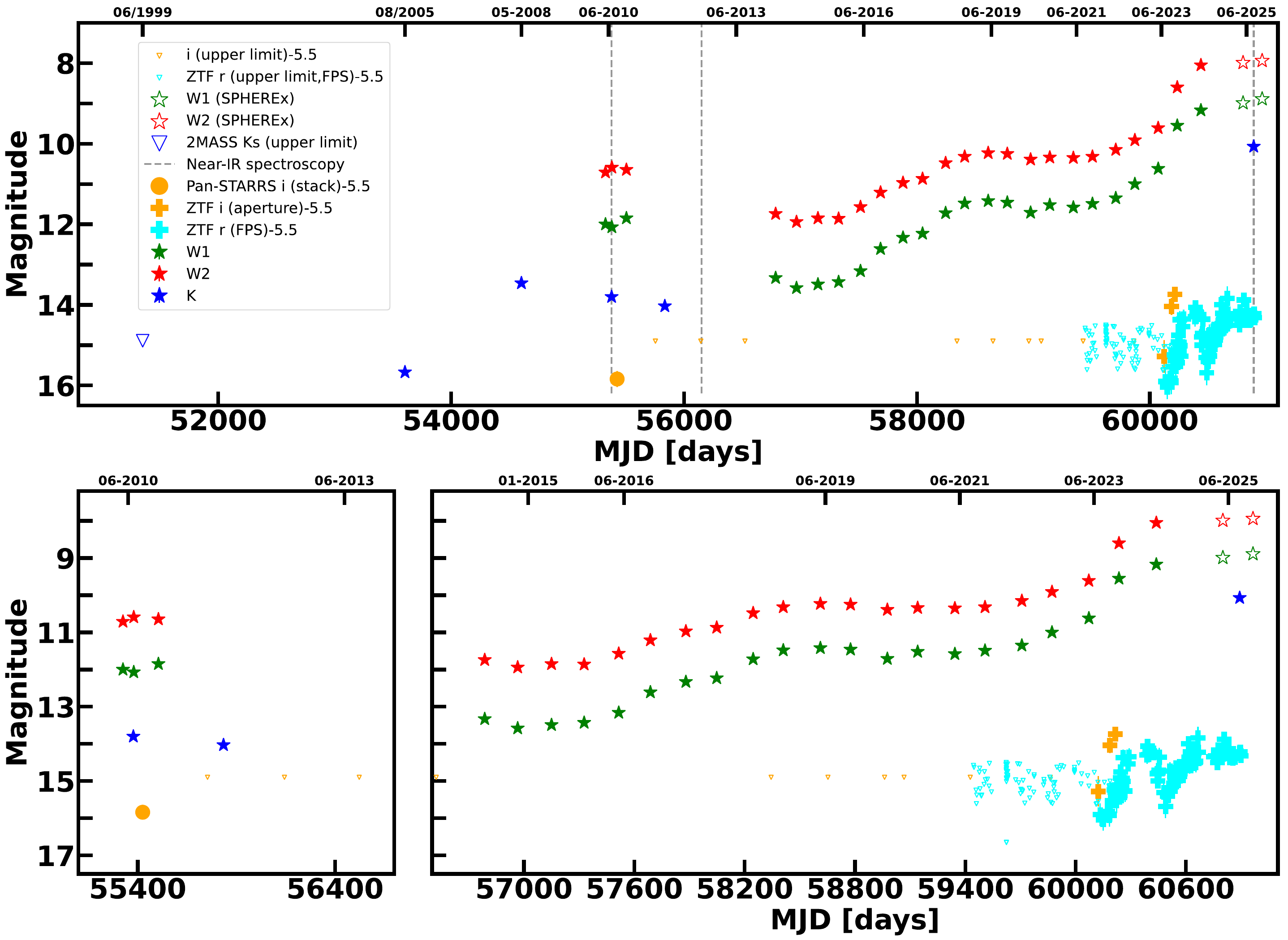}}
	 \caption{(Top) 1999-2026 light curve of GPSV16. (Bottom, left) 2010-2013 section of the light curve. (Bottom, right) 2016-2026 section of the light curve. In all plots, open, upside-down triangles denote upper limits on the photometry from various surveys.}
	 \label{fig:lc}
\end{figure*}

\vspace{0.5cm}
\section{Observations}\label{sec:obs}

We have obtained photometry and spectroscopy of GPSV16 from various sources described below. These data encompass observations from 1999 until 2025. The light curves are shown in Fig.\ \ref{fig:lc}, and photometric data are listed in Table \ref{tab:phot}.

\begin{table}
 \begin{flushleft}

\begin{tabular}{lcccc} 
	\hline
 MJD & Mag & Mag\_error & Filter & Source\\
\hline
60133.37679& 21.40& 0.31 & r & ZTF\\
60149.41521 & 21.54 & 0.30 & r & ZTF\\
60174.38530 & 21.32 & 0.30 & r & ZTF\\
60183.27733 & 21.43 & 0.29 & r & ZTF\\
... &  ... & ... & ... & ... \\
60235.48970 & 9.55 & 0.02 & W1 & NEOWISE\\
60235.54150 & 8.60 & 0.04 & W2 & NEOWISE\\
60438.61190 & 9.17 & 0.03 & W1 & NEOWISE\\
60438.65470 & 8.05 & 0.04 & W2 & NEOWISE\\
60800.60419 & 8.99 & 0.05 & W1 & SPHEREx\\
60800.60419 & 7.99 & 0.05 & W2 & SPHEREx\\
60964.40179 & 8.89 & 0.05 & W1 & SPHEREx\\
60964.40179 & 7.94 & 0.05 & W2 & SPHEREx\\

\hline
	\end{tabular}
    \caption{Optical to mid-IR photometry of GPSV16. The full version of the table is available online.}
\label{tab:phot}
    \end{flushleft}
\end{table}


\subsection{Mid-IR Photometry}\label{ssec:wise}

{\it WISE} surveyed the entire sky in four bands, W1 (3.4 $\mu$m), W2 (4.6 $\mu$m), W3 (12 $\mu$m), and W4 (22 $\mu$m), with spatial resolutions of 6.1\arcsec, 6.4\arcsec, 6.5\arcsec, and 12\arcsec, respectively, from 2010 January to September \citep{2010Wright}. The survey continued as the NEOWISE Post-Cryogenic Mission, using only the W1 and W2 bands, for an additional 4 months \citep{2011Mainzer}. In September 2013, WISE was reactivated as the NEOWISE-reactivation mission \citep[NEOWISE-R,][]{2014Mainzer}. NEOWISE-R was decommissioned in August 2024, and the final data release included observations through 31 July 2024. For each visit to a particular area of the sky, {\it WISE} performed several photometric observations over a period of $\sim$few days. Each area of the sky was observed similarly every $\sim$ 6 months.

\begin{figure}
	\resizebox{\columnwidth}{!}{\includegraphics[angle=0]{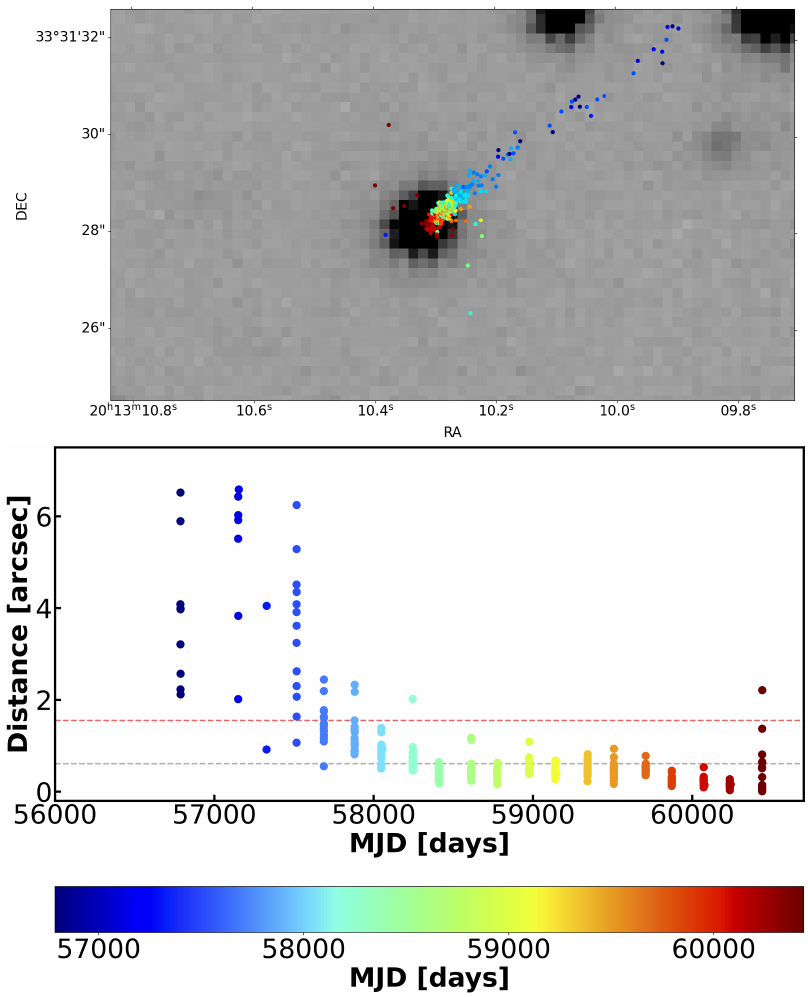}}
	 \caption{(Top) $14\arcsec\times7\arcsec$ UKIRT K band image (observations in 2008, during maximum brightness of the first outburst) centered at the location of GPSV16}. The coloured circles mark the centroids of single-epoch detections from the NEOWISE survey in the range $56800<$MJD$<61000$~d. (Middle) Distance of the centroids of NEOWISE detections towards the coordinates of GPSV16. (Bottom) Diagram showing the colours used to mark NEOWISE detections in the top and middle panels. The colours go from blue at the earlier epochs to red at the latest available data.
	 \label{fig:wise_udet}
\end{figure}

For the analysis of GPSV16, single-epoch data were collected from the NASA/IPAC Infrared Science Archive (IRSA) catalogues within a 3\arcsec\ radius of the YSO coordinates. In addition to the single-epoch data, we also constructed an averaged light curve. Following the procedures described by \citet{2021Park}, we averaged the single-epoch data collected over a few days to produce one epoch of photometry every six months. We noted that the first four epochs of NEOWISE data were missing from the mid-IR light curve. Investigating the NEOWISE detections within 10\arcsec\ of GPSV16 (see Fig.\ \ref{fig:wise_udet}) shows that the first four epochs are unavailable because: 1) the majority of the detections are located beyond 3\arcsec\ from the source, and 2) although some detections fall within 3\arcsec\ the averaging method also makes further cuts to select detections that are within two standard deviations from the mean distance to all detections. Combined with the further requirement that at least three single-epoch detections be required to create one data point every six months, we do not have averaged data during the observations from $56600<MJD<57600$~d in the single-epoch catalogues.  

To obtain the NEOWISE epochs from $MJD<57600$ we queried the unTimely catalogue \citep{2023Meisner}. This catalogue provides time-domain photometry from stacked NEOWISE images, therefore reaching deeper magnitudes. We find that for $MJD<57600$, GPSV16 had faded to $W1\sim13.5$ and $W2\sim11.9$~mag, close to the faint limit of NEOWISE, which probably explains the lack of detections in the single-epoch catalogues.

\begin{figure}
	\resizebox{\columnwidth}{!}{\includegraphics[angle=0]{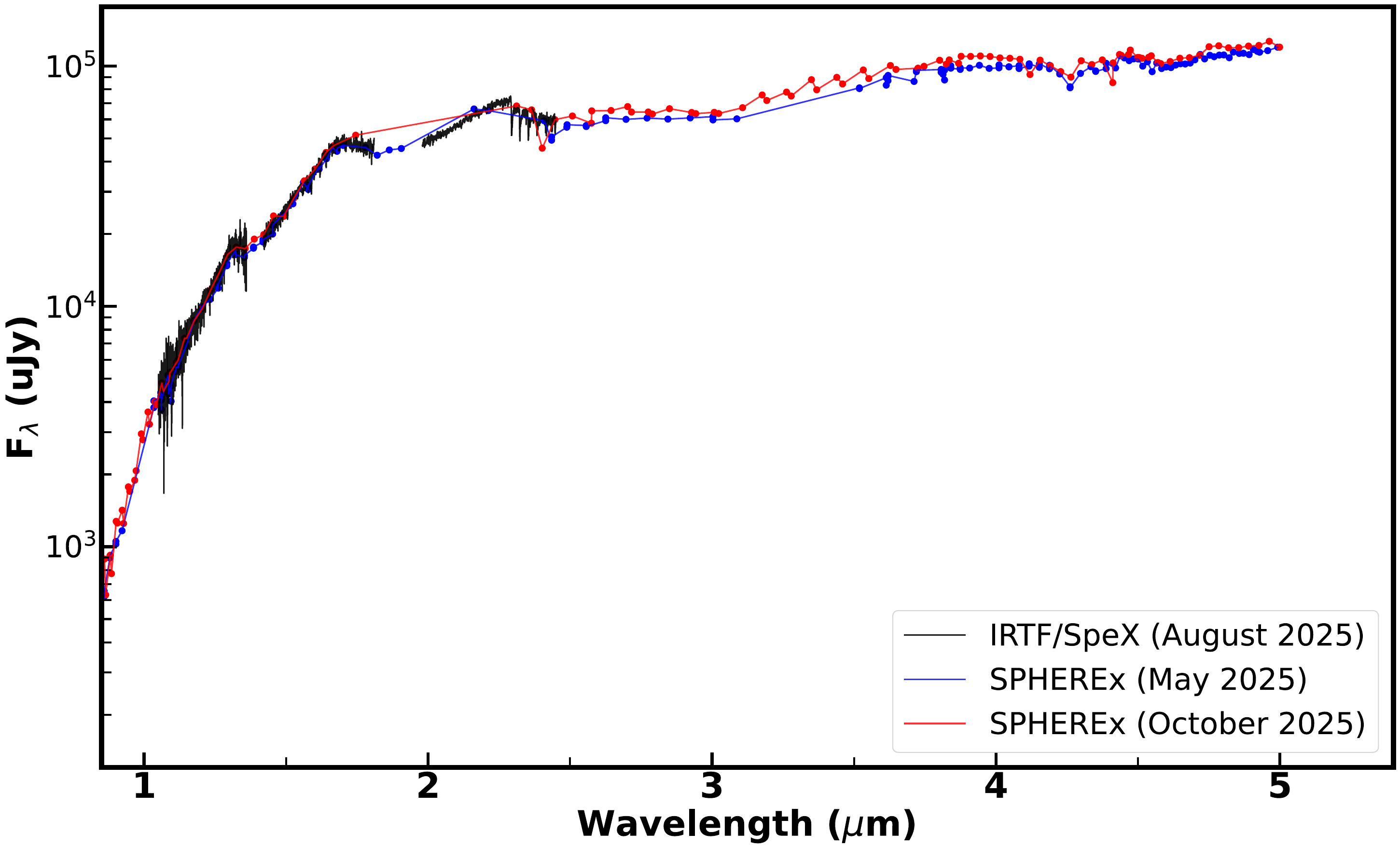}}
	 \caption{SPHEREx spectrum of GPSV16 obtained at two epochs in May 2025 (blue) and October 2025 (red). We show the IRTF/SpeX spectrum obtained in August 2025 for comparison. }
	 \label{fig:spherex}
\end{figure}

The Spectro-Photometer for the History of the Universe, Epoch of Reionization and Ices Explorer (SPHEREx) space mission \citep[][]{2020Crill} is an all-sky survey that provides low resolution ($R=40-130$) $0.7-5 \mu$m spectroscopy every $\sim6$ months. We make use of the online public SPHEREx Spectrophotometry Tool\footnote{https://irsa.ipac.caltech.edu/applications/spherex/tool-spectrophotometry} to obtain SPHEREx spectrophotometry of GPSV16. The query returns a full spectrum for two different epochs at MJD$\sim60800$ and MJD$\sim60964$ (shown in Fig.~\ref{fig:spherex}). We perform synthetic photometry of the SPHEREx spectra at these two epochs, which show that the source is still at the peak brightness of the FUor outburst.

\subsection{Optical and near-IR photometry}\label{ssec:opt}

The UKIDSS GPS survey used the UKIRT telescope to provide near-IR photometry of the Galactic plane accessible from the northern hemisphere. The survey covered the location of GPSV16 in K-band at three epochs: October 2005, May 2008, and October 2011. In the 2008 epoch, the survey also provides J- and H-band observations. An additional K-band epoch in 2010 arises from VLT/ISAAC observations described by \citet{2014Contreras}, while the point in 2025 comes from synthetic photometry of an IRTF/SpeX spectrum (Section \ref{ssec:irtf}). This region of the sky was observed by 2MASS during June 1999, but GPSV16 was not detected, providing an upper limit of $K_{\rm s}=14.3$~mag.

The Panoramic Survey Telescope and Rapid Response System \citep[Pan-STARRS,][]{2016Chambers} and the Zwicky Transient Facility \citep[ZTF, ][]{2019Bellm} provide optical ($g, r, i$) observations of the GPSV16 region between 2010-2012, and from August 2018 onward, respectively. The Pan-STARRS data were collected from the Mikulski Archive for Space Telescopes (MAST), while the latest data release (DR23) of ZTF (observations from 2018 until August 2025) was queried from the IRSA. The source is detected only in one epoch of Pan-STARRS observations at $i=21.4\pm0.2$. 

\begin{figure}
	\resizebox{\columnwidth}{!}{\includegraphics[angle=0]{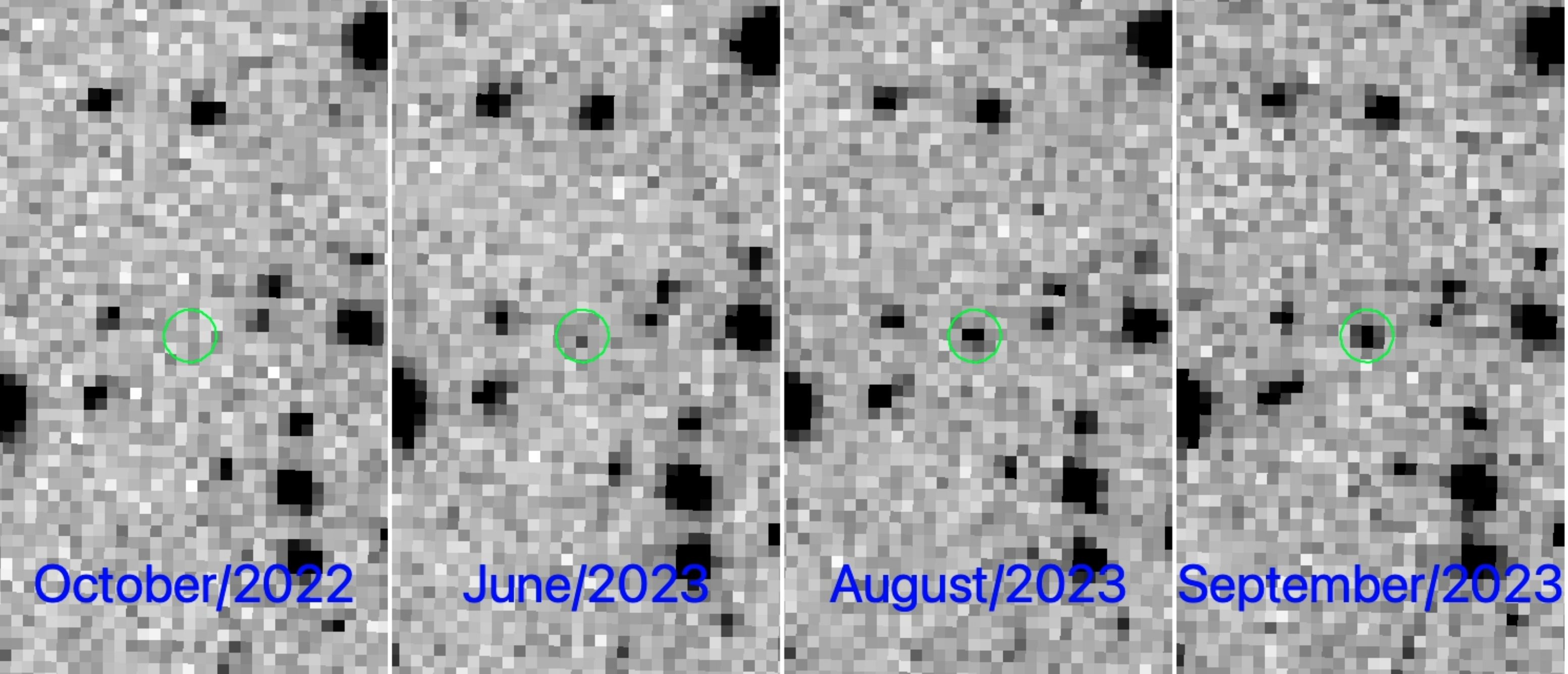}}
	 \caption{$60\arcsec\times30\arcsec$ single epoch i-band images from the ZTF survey at different epochs (marked in the images). The green circles mark the location of GPSV16. In the images North is up and East is to the left.}
	 \label{fig:irise}
\end{figure}

Inspection of single-epoch ZTF images from the IRSA shows that the source becomes visible in $\sim$2024. The single-epoch $i$ images indicate a fast rise. The source is not apparent in the October 2022 and June 2023 images, but it is evident in the August and September 2023 images, with what appears to be a large change between the last two (see Fig.\ \ref{fig:irise}). GPSV16 is observed in some $ r$-band images from March 2024 onward, although it lies very close to the survey's 5$\sigma$ detection limit.

The ZTF forced-photometry service \citep{2019Masci} provides photometry of the public portion of the survey ($g$ and $r$ bands). We obtain photometry in $r$, with the latest data taken in August 2025. To measure the change of the brightness in $i$, we performed aperture photometry for the location GPSV16 (green circle in Fig. \ref{fig:irise}) and of stars in the field around the source. For the photometry, we selected the available i-band images from September 2018 to September 2023. Photometry was obtained with a 2\arcsec\ aperture using standard IRAF routines. We selected 14 stars with low photometric scatter, observed as part of the Pan-STARRS survey \citep{2016Chambers}. The latter are used to correct the apparent magnitude of GPSV16 into the Pan-STARRS system. GPSV16 is only detected in the June, August, and September 2023 images, consistent with the visual inspection from Fig.\ \ref{fig:irise}.

\subsection{IRTF/SpeX Spectroscopy}\label{ssec:irtf}

We obtained a near-IR spectrum of GPSV16  using SpeX \citep{rayner03} mounted at the NASA Infrared Telescope Facility (IRTF) on Mauna Kea (programme 2025B096, PI Contreras Pe\~{n}a). GPSV16 was observed on 5 August 2025 (HST). The cross-dispersed spectra cover 0.7--2.5 $\mu$m  spectra at $R=1200$, obtained with the 0.5\arcsec\ slit. The total integration time was 360s with individual exposures of 60s.  The bright A0V standard star HD171149 was also observed for telluric calibration. 

For purposes of measuring the visual extinction towards GPSV16, we obtained the publicly available SpeX spectrum of FU Ori. The observations carried out on 27 January 2020 were part of programme 2019B073 (PI Connelley) and were acquired via the The IRTF Data Archive at IRSA. All spectra were reduced and calibrated using Spextool version 4.1 \citep{cushing04}.

The flux calibration of SpeX data is generally robust \citep{2003Vacca}. The comparison between the IRTF/SpeX and SPHEREx spectra of GPSV16 (Fig. \ref{fig:spherex}) confirms the accuracy of the SpeX flux calibration. We calculated synthetic photometry from the SpeX spectra of GPSV16 using the 2MASS JHK filter profiles. The photometric data are presented in Table \ref{tab:phot} and Fig.\ \ref{fig:lc}.


\section{Association with a star forming region}\label{sec:sfr}

GPSV16 is located 170\arcsec\ from the HII region G71.52$-$00.38 \citep{2016Xu, 2019Reid}. The distance to this region is estimated at $3.61^{+0.18}_{-0.16}$~kpc \citep[obtained from VLBI trigonometric parallax and proper motion observations by][]{2016Xu}. To our knowledge, there are no detailed studies of the YSO population towards the region. 

We use near-IR JHK observations from the UKIDSS GPS survey to search for sources exhibiting near-IR excess within a region within 5\arcmin\ of G71.52$-$00.38. We queried the Wide Field Camera (WFCAM) Science Archive
\citep[WSA,][]{2008Hambly} to select sources classified as stellar (objects with image profile classifier $mergedClass=-1$) and that were detected in the three filters.

\begin{figure*}
\centering
	\resizebox{\columnwidth}{!}{\includegraphics[angle=0]{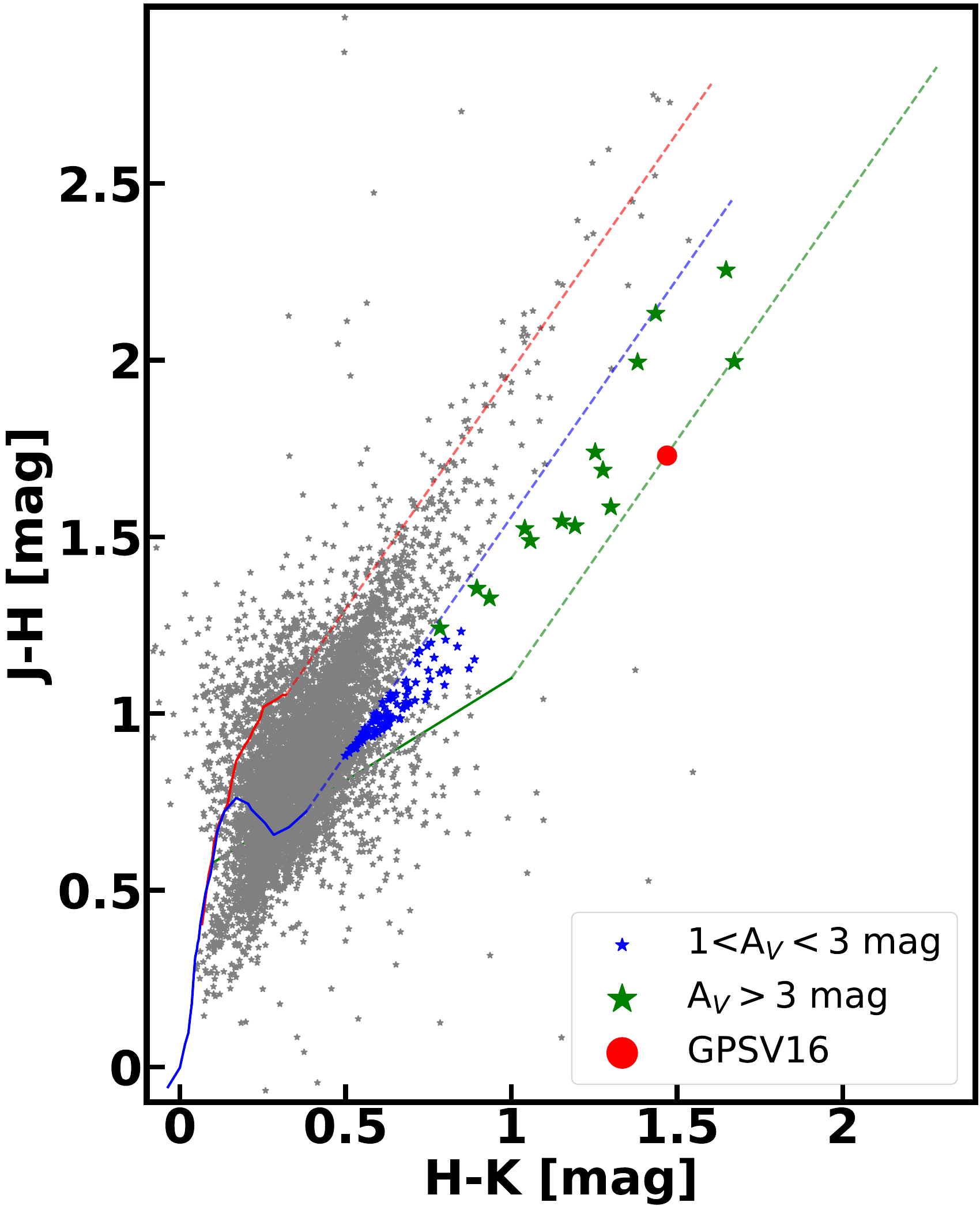}}
    \resizebox{\columnwidth}{!}{\includegraphics[angle=0]{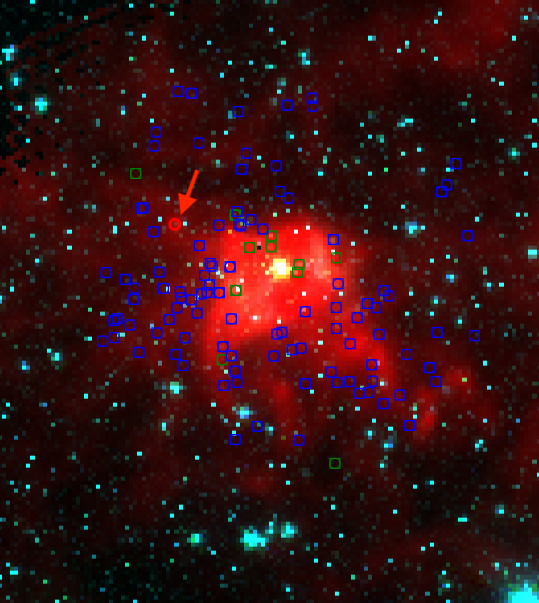}}
	 \caption{(Left) J-H versus H-K$_{\rm s}$ colour-colour diagram of UKIDSS GPS sources (grey circles) located within 5\arcmin\ of HII region G71.52$-$00.38. The classical T Tauri locus of \citet{1997Meyer} is presented (solid green line) along with intrinsic colours of dwarfs and giants (blue and red solid lines, respectively) from \citet{1988Bessell}. Reddening vectors of A$_{V}=20$~mag are shown as dotted lines. Stars to the right of the reddened dwarf colours and above the CTTS locus are marked as potential members of the SFR. Sources in this region with $1<$A$_{V}<3$ are marked by blue stars, while sources with A$_{V}>3$~mag are shown as green stars. The location of GPSV16 is marked by the red circle. (Right) $14\arcmin\times14\arcmin$ false colour image of the HII region G71.52$-$00.38 using images from {\it Spitzer} 3.6 $\mu$m (blue), 4.5 $\mu$m (green) and {\it Herschel} 250 $\mu$m (red). The location of stars that are selected as potential members of the SFR from their near-IR colours is marked using the same colour scheme as in the {\it left} figure.  The location of GPSV16 is marked by the red cicle and arrow.}
	 \label{fig:cmd}
\end{figure*}

Figure \ref{fig:cmd} shows the $J-H$ versus $H-K$ color-color diagram of the selected sources. We select objects with near-IR excess (i.e.,\ candidate YSOs) as sources with $H-K$ colours that are larger than those of reddened dwarf and giant stars. The selected sources follow the (reddened) loci of classical T Tauri stars derived by \citet{1997Meyer}. The selected sources include the JHK photometry of GPSV16 during the first outburst (observations from 2008). 

Figure \ref{fig:cmd} shows the location of the candidate YSOs. The figure shows that these are locally clustered around the HII region. The existence of a population of candidate YSOs around G71.52$-$00.38, and the fact that GPSV16 is one of the sources, makes it very likely that this eruptive YSO is associated with the star-forming region. We therefore assume a distance of 3.61 kpc towards GPSV16.

\section{The outbursts of GPSV16}\label{sec:outbursts}

The long-term light curve of GPSV16 (Fig.\ \ref{fig:lc}) illustrates the complexity of its variability. Here, we discuss the properties of the two high-amplitude outbursts of the YSO.

\subsection{The first outburst: 2005-2012} 

GPSV16 showed high amplitude variability, with $\Delta {\rm K}_{\rm s}=2.2$~mag between two epochs observed in 2005 and 2008 within the UKIDSS GPS survey \citep{2014Contreras,2017Lucas}. The lack of detection by 2MASS \citep{2006Skrutskie} observations in 1999 (with an upper limit of $K_{\rm s}=14.6$~mag), supports the scenario of an outburst between 2005 and 2008. The source decreased in brightness since 2008, but remained close to the peak in 2010 \citep[VLT/ISAAC K-band photometry from][and Spitzer and WISE mid-IR photometry]{2014Contreras} and  2011 (additional K-band epoch from GPS) observations.

During this first outburst, GPSV16 shows strong Br$\gamma$ (2.16$\mu$m) and $^{12}$CO (2.29$\mu$m) emission, apparent in both VLT/ISAAC and Gemini/NIFS spectroscopy \citep[see Figure 10 in][]{2014Contreras}. These features, along with the apparent years-long duration of the outburst, lead to an intermediate classification between EX Lupi-type and FUor eruptive YSO, or PVM, for GPSV16.   

The first four epochs of $3-5 \mu$m photometry from NEOWISE suggest that GPSV16 became fainter at some point between 2012 and 2013 (see Section \ref{ssec:wise}). The K-band data also show that GPSV16 is fading after the 2008 peak brightness observation. The region containing GPSV16 was also covered by the Pan-STARSS survey between 2009 and 2012. However, the source is detected only in one epoch at $i=21.4\pm0.2$.


\subsection{The second outburst: since 2016}\label{ssec:sout}

The brightness of GPSV16 has been increasing since 2016 (if not earlier). The changes in the W1 and W2 amplitudes between the June 2016 and May 2024 epochs are 3.4 and 3.2 magnitudes, respectively. The time-series photometry indicates that this was a two-stage rise (see Fig.\ \ref{fig:lc}). A slow rise of $\sim$1 mag over $\sim$2.5 years was followed by a rise with a larger amplitude ($\sim$2.4 mag) and a similar timescale of 2.5 years. A two-stage mid-IR rise was also observed in the outburst of Gaia17bpi \citep[see][]{2018Hillenbrand}. A more thorough analysis and discussion on these types of light curves is presented below, in Section \ref{ssec:irprec}.

\begin{figure}
\centering
\resizebox{\columnwidth}{!}{\includegraphics[angle=0]{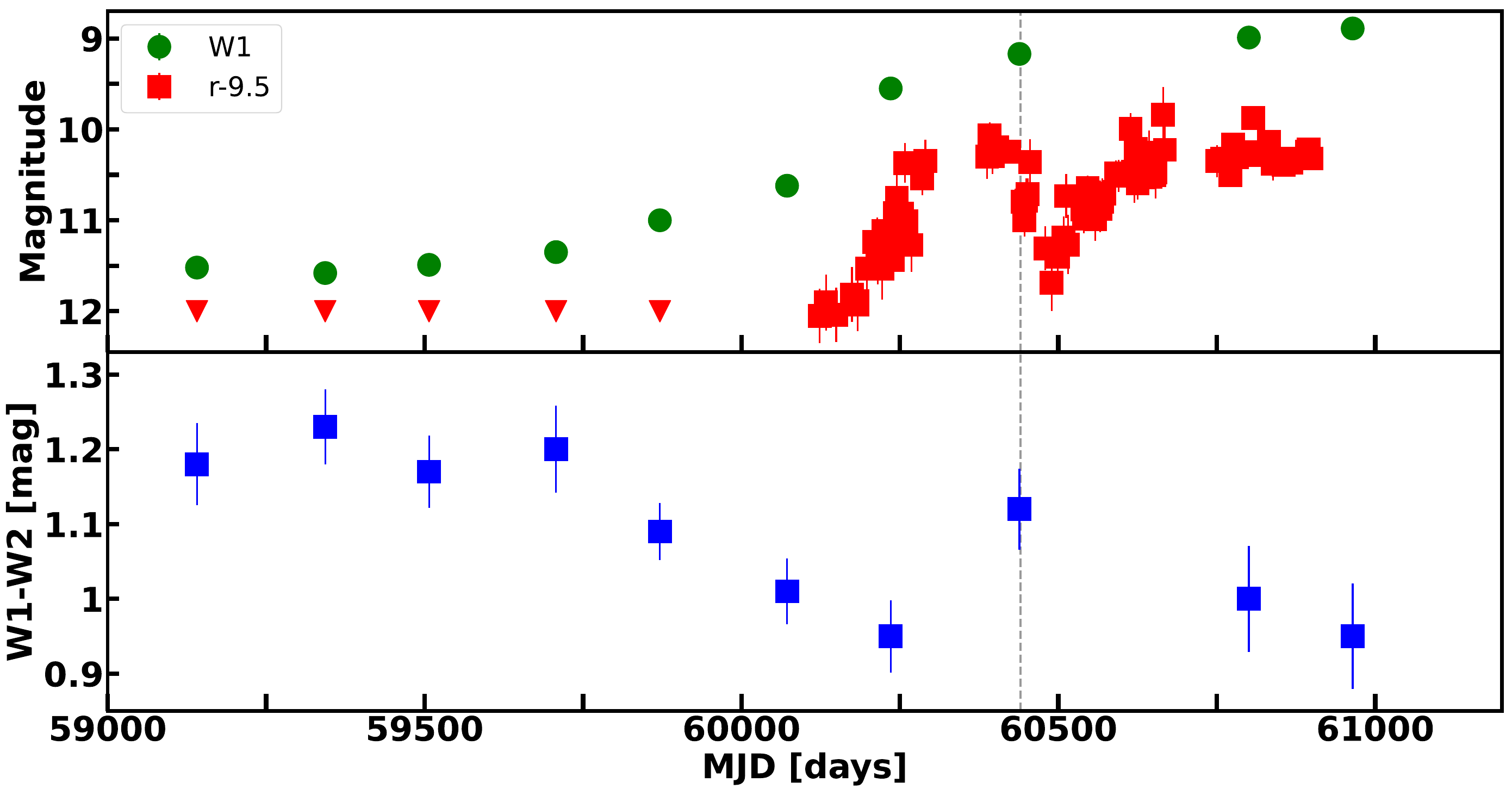}}
	 \caption{(Top) ZTF r and WISE $W1$ light curve of GPSV16 during $59000<$MJD$<61200$~d. (Bottom) $W1-W2$ ligth curve of the source during the same period of time. The gray line in both plots marks the approximate start of the optical fading.}
	 \label{fig:rstruct}
\end{figure}

The ZTF r photometry shows that the optical brightening follows the second-stage of the mid-IR change. The true amplitude of the optical outburst cannot be quantified because the source is fainter than the ZTF detection limit ($r\sim21.5$) for most epochs. During the peak in $r$, the source is $\sim2$ magnitudes brighter than the faint limit. A rapid rise is also observed in the $i$ images of the ZTF survey.

The ZTF r light curve becomes fainter by $\sim1$~mag, with the dip starting at MJD$\sim60430$~d. The $W1-W2$ colour was becoming bluer as the optical outburst was getting brighter, but at around the epoch of the optical fading event, GPSV16 becomes brighter ($\Delta W1=0.5$~mag) and redder (by $\Delta(W1-W2)=0.2$~mag) in the mid-IR (see Fig. \ref{fig:rstruct}). A brighter mid-IR emission and fainter optical emission may be explained by an increased scale height of the inner disk, which can lead to higher optical extinction and larger surface area of warm dust (\citealt{2019Bryan,2021Covey}). A similar type of variability has been observed in the recent outburst of PR Ori B (Contreras Pe\~{n}a et al., in prep) which may suggest that this is a common feature in FUor outbursts.

The synthetic photometry from the 2025 SpeX spectrum yields $K_{\rm s}=10.1\pm0.02$ (the uncertainty given by the signal-to-noise ratio in the K-band). Assuming that the 2005 GPS epoch corresponds to the quiescent magnitude of GPSV16, the amplitude of the outburst is $\Delta K=5.6\pm0.1$~mag.

\subsection{Near-IR spectroscopy of the second outburst}\label{sec:irspec}

\begin{figure*}
\centering
\resizebox{1.9\columnwidth}{!}{\includegraphics[angle=0]{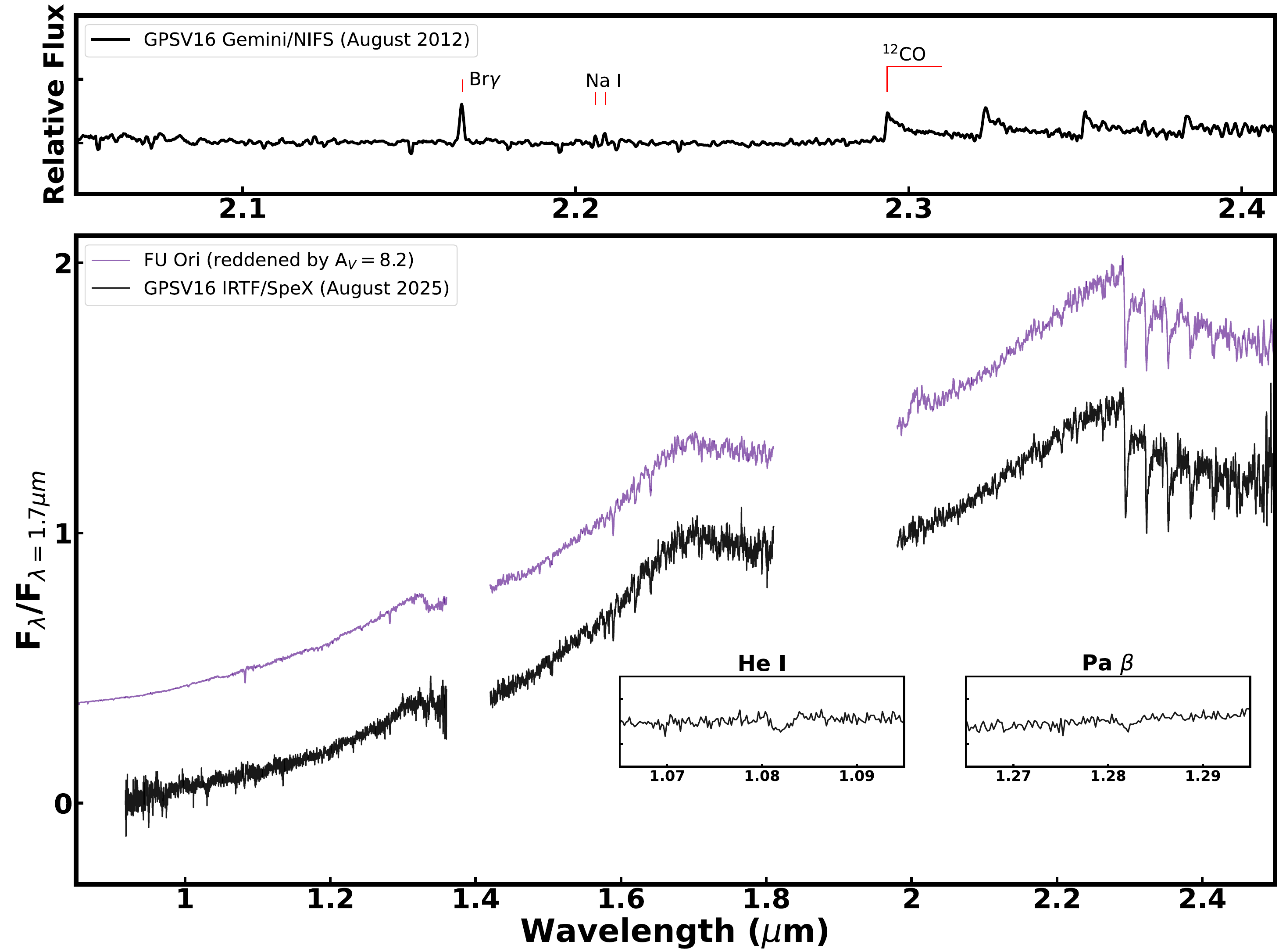}}
	 \caption{(Top) $2-2.5 \mu$m Gemini/NIFS spectrum of GPSV16 taken during the first outburst (August 2012). (Bottom) Spectrum of GPSV16 (black) between 0.8 and 2.5 $\mu$m taken during the second outburst (August 2025). The spectrum of FU Ori Ori (purple) is shown for comparison, and it has been reddened by A$_{V}=8.2$~mag to match the spectrum of GPSV16. The insets show the Paschen $\beta$ and \ion{He}{1} lines of GPSV16.}
	 \label{fig:jspec}
\end{figure*}

The $0.8-2.5 \mu$m SpeX spectrum (Fig.\ \ref{fig:jspec}) of GPSV16 taken during the second outburst shows a clear change from the spectrum taken during the first outburst. In 2025, the source no longer displays emission lines in the K-band. Instead, it shows strong $^{12}$CO ($\Delta\nu=2$) ro-vibrational absorption bands, along with weak absorption from Na I (2.21$\mu$m) and Ca I (2.26$\mu$m). These features are more consistent with absorption in the low-gravity atmospheres of M-type giant stars and are defining characteristics of the near-IR spectra of FUors \citep{2018Connelley}. 


The H-band spectrum of GPSV16 shows a triangular shape arising from absorption from water vapor bands, also a defining characteristic of FUors. This feature correlates with the ratio of the viscous disk's luminosity to that of the stellar photosphere, indicating a strong accretion outburst in FUors \citep{2022Liu}.

The low SNR in the spectrum at wavelengths shorter than 1$\mu$m (SNR$<3$) prevents us from detecting features commonly associated with outflows in FUors, such as the Ca II triplet (0.850, 0.854, and 0.866 $\mu$m) and the O I (0.77 and 0.84 $\mu$m) lines. At 1.083 $\mu$m, we identify the Paschen $\beta$ and \ion{He}{1} lines in absorption. The \ion{He}{1} line is associated with outflows driven by accretion outbursts and is commonly observed as an absorption feature in FUors \citep{2018Connelley,2023Ghosh}. 

In Fig.\ \ref{fig:jspec} we compare the spectra of GPSV16 to that of the archetype of the FUor class, FU Ori. These objects show remarkable similarities. All the evidence from the second outburst points to GPSV16 being a new, and ongoing, FUor.

\section{Properties of the second outburst}\label{sec:secout}

\subsection{The spectroscopic change}

To the best of our knowledge, GPSV16 is the first eruptive YSO that has shown such a marked change of spectroscopic characteristics between outbursts. The recently discovered eruptive YSO TCP J21220926$+$4926242 is undergoing a high amplitude outburst with FUor spectroscopic characteristics \citep{2026Burlak}. \citet{2025Stecklum} shows that the object displayed previous variability that resembles that of EX Lupi-type objects, although no spectroscopic evidence for that classification is presented. V1647 Ori showed an emission line-dominated spectrum in its 2004-2006 outburst, while a second outburst, begun in 2011, displays the triangular H-band shape due to water vapor absorption. However, $^{12}$CO at 2.29 $\mu$m was not detected in emission or absorption.

In the case of GPSV16, there is a distinct difference in the amplitudes of the two outbursts, with $\Delta K_{\rm s}=2.2$~mag and $\Delta K_{\rm s}=5.6$~mag for the first (2005-2012) and second (since 2016) outbursts, respectively. The difference in amplitudes can be explained by the different values of the accretion rate reached during the two outbursts \citep[see e.g.][]{2022Liu}. The accretion rate also influences the characteristics observed in the near-IR spectra during an outburst.  

As the outburst progresses and the accretion rate increases, the spectrum goes from being dominated by features from the stellar photosphere and the passively heated disk, at lower accretion rates, to a FUor spectrum as the accretion rate reaches a few$\times10^{-5}$~M$_{\odot}$~yr$^{-1}$. At these high accretion rates, the viscously heated disk dominates the emission from the system \citep{2022Liu,2022Rodriguez,2024Carvalho}. In their models, \citet{2022Liu} define a parameter, $\eta$, as the ratio of the summed H- and K-band fluxes of the viscous disk to that of the photosphere. Therefore, $\eta=1$, implies equal contribution, while $\eta=5$ defines the point of sufficient dominance from the viscous disk such that the spectrum of a YSO should resemble that of FUors.

\begin{figure}
\centering
	\resizebox{\columnwidth}{!}{\includegraphics[angle=0]{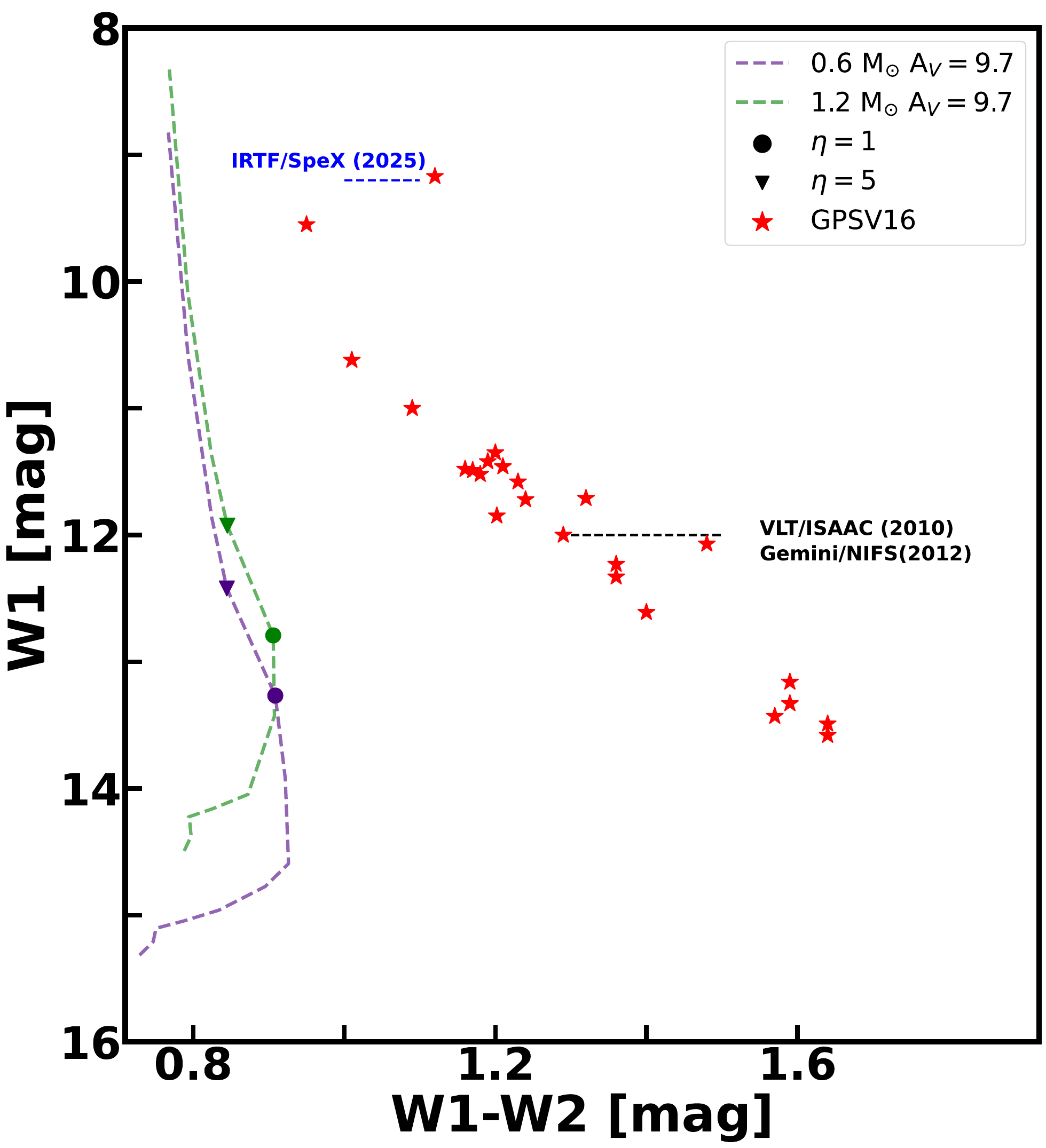}}
	 \caption{W1 versus W1$-$W2 for the light curve of GPSV16 (red stars). In the figure, we compare with isomass curves for an FUor outburst \citep{2022Liu} at a distance of 3.61 kpc (Section \ref{sec:sfr}) with the mass of the central star, M${_{\ast}}=0.6$~and $1.2{\rm M}_{\odot}$, and using A$_{V}=9.7$~mag. The solid symbols mark discrete values of $\dot{\rm M}$ used in the models of \citep{2022Liu}, where $\eta=1$ (circles) and $\eta=5$ (upside down triangle). Finally, we also mark the approximate magnitude and colours of GPSV16 during the spectroscopic follow-ups of the first (black dotted line) and the second (blue dotted line) outbursts. }
	 \label{fig:isomass}
\end{figure}


In Fig.\ \ref{fig:isomass} we compare the evolution of the outburst of GPSV16 in the W1 versus W1$-$W2 color-magnitude diagram, with isomass curves for an FUor outburst from \citet{2022Liu}, set at the distance (3.61 kpc) and extinction (A$_{V}=9.7$~mag) of our YSO. We note that GPSV16 shows redder colours than the models, similar to some of the known FUors \citep[see the discussion by][]{2022Liu}. The redder colours may be an indication that the disk in GPSV16 is not in a steady state, but instead has a larger accretion rate at larger radii in the disk. This would also explain the lag between the IR and optical outburst (Section \ref{ssec:irprec}). The redder colours, however, persist even at the brightest point of the mid-IR CMD, when the disk should have reached a steady state, indicating that an additional source of emission, such as from a circumstellar envelope, must exist \citep{2022Liu}.

The comparison with the \citet{2022Liu} models is still useful, as it indicates that during the second outburst, the accretion rate is large enough ($\eta>>5$) for the viscous disk to completely dominate the emission of the system, leading to the FUor spectrum.


During the first outburst, the magnitude falls in a region where models are close to $\eta=5$ of the isomass curves. At these accretion rates, we would expect to observe features arising from the viscous disk in the spectra. However, GPSV16 displays $^{12}$CO emission, which is usually associated with the magnetospheric accretion still controlling the accretion process. \citet{2023Wang} also find that the 2008 outburst in EX Lupi reached an accretion rate where the viscous disk should be sufficiently dominant for its spectrum to display FUor characteristics \citep[see also][]{2023Cruz}. \citet{2023Wang} explain the lack of these features by the innermost disk not being fed at a sufficient rate to heat the disk sufficiently to dominate the spectrum. The models by \citet{2022Liu} are based on simplifying assumptions for the geometry and strength of the magnetic field. Differences in these parameters would affect the spectroscopic characteristics of an outburst when reaching the transition region at $1<\eta<5$.



\subsection{Luminosity of the second outburst}

We estimate the luminosity of the FUor outburst using the bolometric corrections provided by \citet{2024Carvalho}. For the calculation, we assume a distance of 3.61 kpc to GPSV16 (see Section 
\ref{sec:sfr}).

We derive an approximate value of the reddening, A$_{V}$, from near-IR photometry. The GPS observations during the first outburst yield $J-K_{\rm s}=3.2\pm0.03$~mag. From the synthetic photometry of the IRTF/SpeX spectrum, we derive $J-K_{\rm s}=2.7\pm0.14$~mag. These values are 1.84 and 1.34 magnitudes redder than the $J-K_{\rm s}$ color of FU Ori. Using the color excess and the extinction law of \citet{2021Gordon}, we estimate the reddening towards GPSV16 as A$_{V}=({\rm excess}/0.165)+{\rm A}_{V}{\rm (FU~Ori)}$. Taking ${\rm A}_{V}=1.5\pm0.2$~mag for FU Ori \citep{2018Connelley}, we determine A$_{V}=12.6\pm0.3$~mag and A$_{V}=9.6\pm0.9$~mag from the GPS and SpeX photometry, respectively. Given that GPSV16 does not resemble a FUor during the GPS observations, the A$_{V}$ determined from these data is likely inaccurate, and we only use it as an upper limit. Finally, to match the spectrum of FU Ori \citep[A$_{V}=1.5\pm0.2$,][]{2018Connelley} with GPSV16 (see Fig.\ \ref{fig:jspec}), we reddened the former by A$_{V}=8.2$~mag, which yields A$_{V}=9.7$~mag for GPSV16, close to the value estimated from photometry. For the remainder of this work, we assume A$_{V}=9.7\pm0.2$~mag for GPSV16.


\begin{figure}
\centering
\resizebox{\columnwidth}{!}{\includegraphics[angle=0]{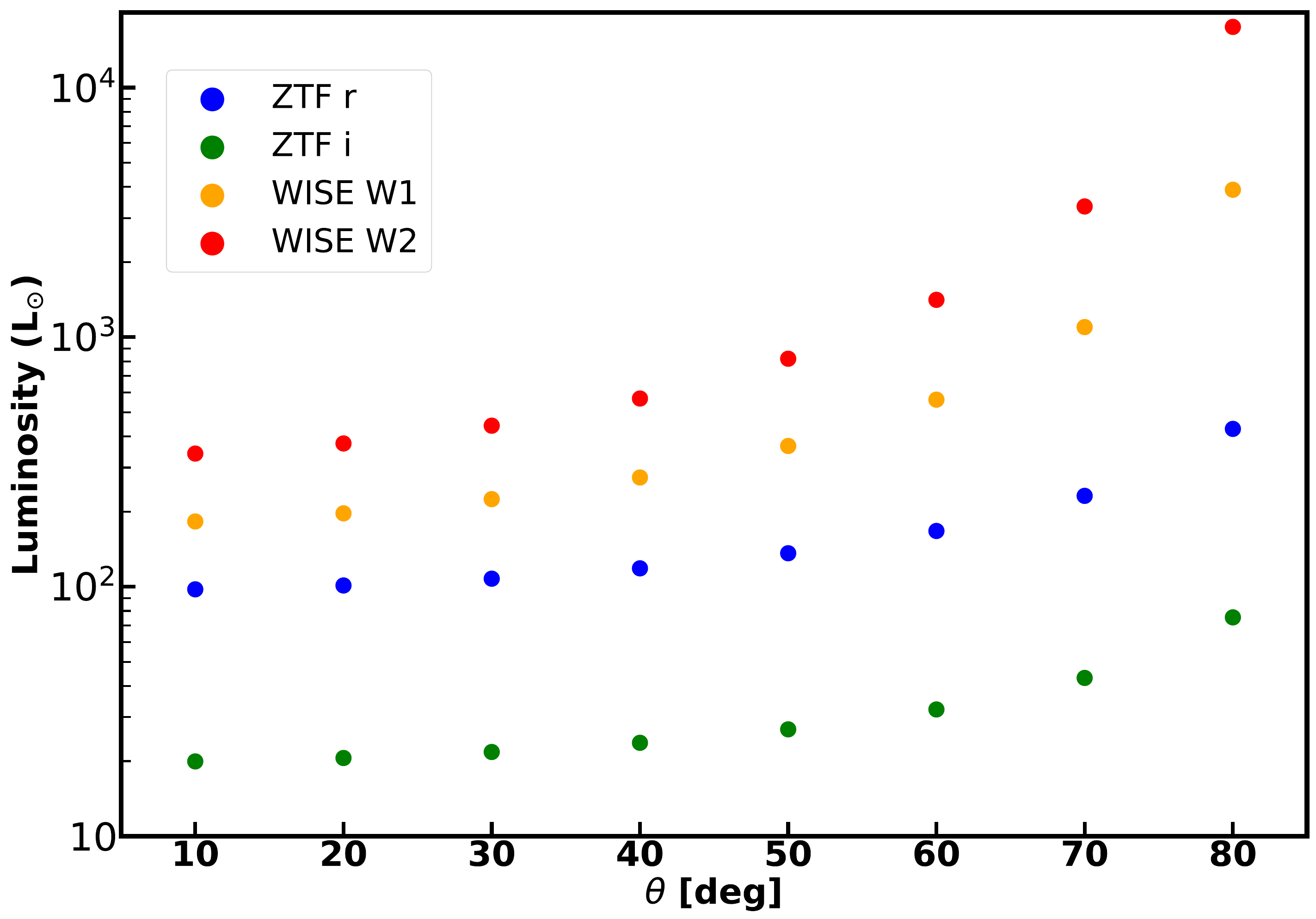}}
	 \caption{Accretion luminosity of the second outburst (since 2016) of GPSV16, derived using the bolometric corrections from \citet{2024Carvalho}. The values are estimated for different inclination angles $10\leq \theta\leq80$~deg from the brightest magnitudes in the filters $r$ (blue circles) and $i$ (green circles) from the ZTF survey, and in W1 (yellow circles) and W2 (red circles) from NEOWISE.}
	 \label{fig:luminosity}
\end{figure}

The values of the luminosity from \citet{2024Carvalho} depend on the inclination angle of the disk, $\theta$. We study the effects of varying the value of $\theta$ in our estimates of the luminosity, L$_{acc}$. We also derive the outburst luminosity from WISE W1 and W2 and ZTF $r$ and $i$ filters. We derive absolute luminosities of the disk using the polynomial coefficients from Table 1 of \citet{2024Carvalho}. Figure \ref{fig:luminosity} shows the luminosity estimated from different filters and varying the inclination angle between 10 and 80 degrees. There is a large spread in the values of L$_{acc}$ with the values derived from the WISE W1 and W2 filters being systematically larger (reaching $10^3$-$10^4$L$_\odot$ for large inclination angles). The mid-IR estimates could be affected by the emission reprocessed by a dust disk component, and could lead to overestimating the luminosity derived at these wavelengths by factors of 2-3 \citep[see the case of V960 Mon,][]{2024Carvalho}.



There are also large differences between the values obtained from optical data. However, it is not clear that the available photometry from the ZTF $i$ filter covers the peak of the outburst. The ZTF $r$ filter covers a larger portion of the outburst and it is clear that it includes its peak. Given this, we favour the accretion luminosity derived from the ZTF $r$ data. The luminosities for the outburst of GPSV16 vary between 97 and 430 L$\odot$, with a median value of L$_{acc}=130$~L$\odot$



\subsection{The two-stage mid-IR brightening}\label{ssec:irprec}

We follow  \citet{2018Hillenbrand} and provide a quantitative analysis of the mid-IR light curve of the second outburst of GPSV16 by fitting a sigmoid function described by $L/(1+e^{-k\times(t-t_{0})})$. Here, $L$ is the amplitude of the outburst and $1/k$ represents the e-folding timescale of the rise. In the fits, $t=0$ is defined as the epoch at which the outburst reaches its maximum brightness, and at $t=t{0}$ the function reaches $L/2$. The latter definition yields a negative $t_{0}$, and we can take $2\times t_{0}$ as the approximate time at which the outburst began.

\begin{figure}
\centering
\resizebox{\columnwidth}{!}{\includegraphics[angle=0]{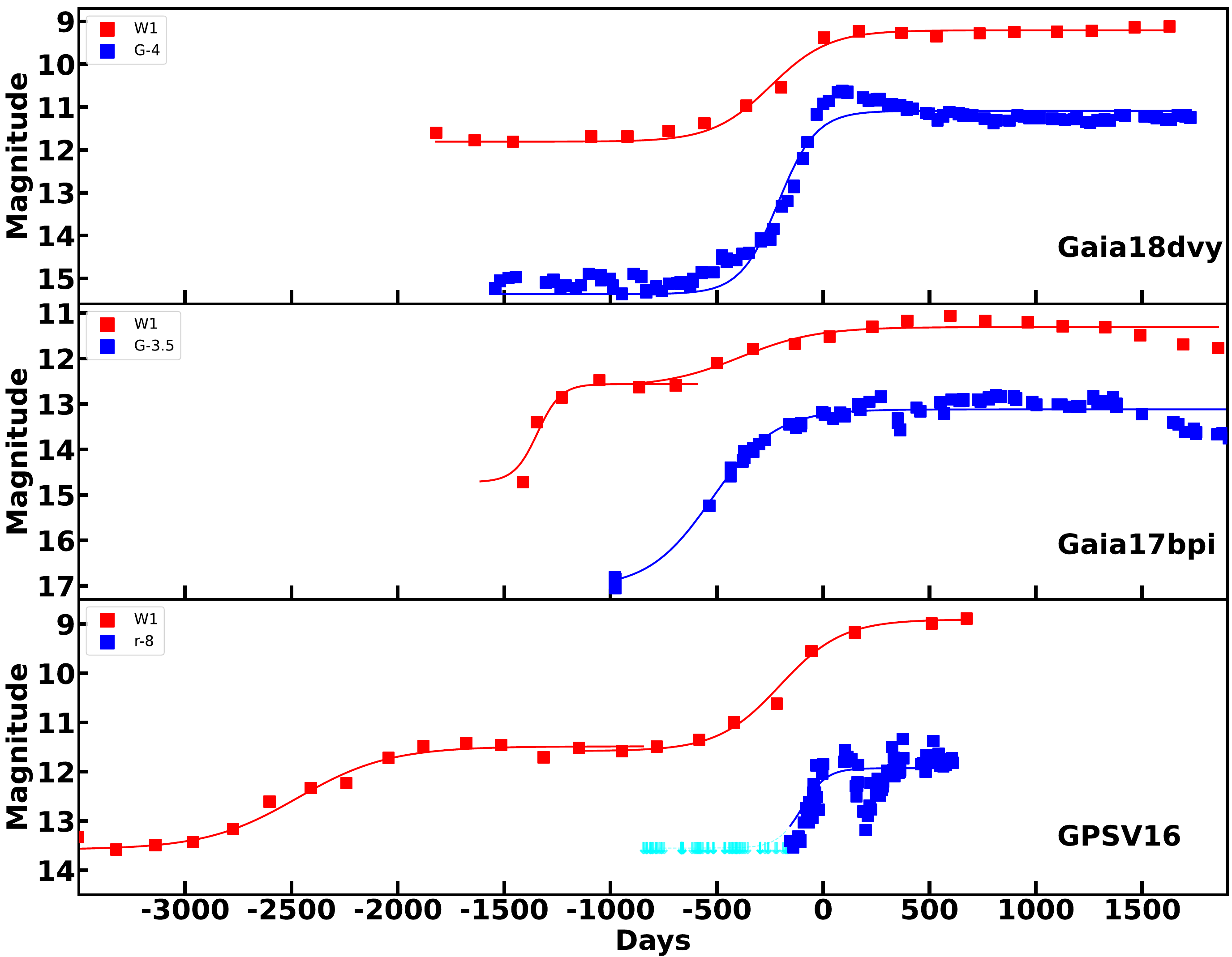}}
	 \caption{Optical (blue) and mid-IR (red) light curves of FUors Gaia18dvy (top), Gaia17bpi (middle), and GPSV16 (bottom). The sigmoid functions fitted to each light curve are shown with solid lines. Upper limits of the r-band photometry of GPSV16 are marked with upside-down, cyan triangles.}
	 \label{fig:comp}
\end{figure}

The second outburst of GPSV16 shows a two-stage rise in the mid-IR (Section \ref{ssec:sout}). Therefore, we fit the two stages separately (the fits to the W1 light curve are shown in Fig.\ \ref{fig:comp}). In the first rise we set $t=0$ at MJD$=58410$~d, and the light curve is fitted with $L=2.09\pm0.14$, $k=0.0046534\pm0.00093053$~d$^{-1}$ and $t_{0}=-591.65\pm84.61$~d. For the second rise we set $t=0$ at MJD$=60290$~d, and the fit yields $L=2.67\pm0.11$, $k=0.0068438\pm0.0007970$ and $t_{0}=-202.23\pm22.68$. The e-folding timescales are 96 and 149 days for the first and second rises, respectively.

The first rise starts 1183 days (at MJD$\sim$57227~d) before reaching a maximum amplitude of 2.1 magnitudes, while the second rise starts 404 days (at MJD$\sim$58886~d) before reaching its maximum amplitude of 2.7 magnitudes. The combination of both fits indicates that the second outburst started $\sim$3063 days (about 8.4 years) before reaching its maximum brightness, with a total amplitude of 4.8 magnitudes. 

The optical outburst appears to follow the second mid-IR rise. We perform a fit to the r-band light curve of the source (see Fig. \ref{fig:comp}). The source is not detected for MJD$<60130$~d, and the data provide only an upper limit. The fit, which uses the upper limits, and where we set $t=0$ at MJD$=60290$~d (same as the second mid-IR rise), indicates $L=1.62\pm0.03$, $k=0.017387\pm0.0033033$ and $t_{0}=-100.0\pm10.7$. The value of $t_{0}$ implies that the optical outburst started 200 days before reaching its peak (at MJD$=60090$~d), about 204 days after the start of the second mid-IR rise. However, because upper limits are used, this fit may not yield accurate properties of the optical outburst.


In Fig.\ \ref{fig:comp}, we also show the fits to the optical and mid-IR light curves of Gaia18dvy and Gaia17pi. The mid-IR brightening also precedes the optical outburst in these two YSOs \citep{2018Hillenbrand,2020Szegedi-Elek}. Using the values of t$_{0}$ in the fits, we estimate that in Gaia18dvy, the lag time between the mid-IR and optical is $\sim72$ days, while in Gaia17bpi the mid-IR brightening starts $\sim$588 days prior to the onset of the optical outburst. 

The outbursts of GPSV16 and Gaia 17bpi show the strongest similarities: both exhibit a long lag time between mid-IR and optical outbursts and two-stage mid-IR brightening. In GPSV16, however, the first stage of mid-IR brightening started at least 8 years before the onset of the optical outburst (if we assume that the onset of the optical outburst occurs during the second rise of the mid-IR light curve). The luminosity of the outburst, $\sim130$ L$_{\odot}$, is also larger than in Gaia17bpi \citep[$\sim8$\,L$_{\odot}$][]{2018Hillenbrand}. This fits well with the models of \citet{2023Cleaver}. In that work, the differences in lag times are attributed to the distance from which the outburst started in the disk. Longer lag times and larger accretion rates occur for outbursts starting at large distances ($\sim1$~AU) in the disk. 

Using a lag time of $\sim8$~years, we can provide a rough estimate of the disk location at the time the outburst started in GPSV16. In $\alpha$-disk models \citep[e.g.][]{1994Bell} the propagation timescale for the accretion wave is given by $t_{prop}=r/(\alpha c_{s})$, with $r$ as the radius where the outburst started, $\alpha$ the viscosity parameter and $c_{s}$ the sound speed. By assuming $t_{prop}=8$~yr, $\alpha=0.1$ \citep[the estimate of $\alpha$ for the outburst of Gaia17bpi in][]{2023Cleaver}, and $c_{s}=2.3$~km s$^{-1}$ for $T=1500$~K, we determine $r\sim0.4$~AU. The latter is larger than the distance of $\sim0.05-0.1$~AU determined for Gaia17bpi \citep{2023Cleaver}.

\section{Summary}\label{sec:summary}

In this work, we present an analysis of a new outburst in the Class I YSO GPSV16. We find that the source is likely associated with the HII region G71.52$-$00.38 and therefore at a distance of 3.61 kpc. The source had already been classified as an eruptive variable YSO due to a $\Delta K=2.2$~mag outburst that lasted longer than two years. In the first outburst, the near-IR spectrum was dominated by emission lines of Na I, H I, and $^{12}$CO \citep{2014Contreras}. The spectroscopic and photometric characteristics of the outburst placed it in the PVM category in \citet{2025Contreras_b}.

The second outburst of GPSV16 started around 2015 with a 2.1 magnitude rise at mid-IR wavelengths. The rise started 1180 days before reaching {a bright plateau}, and stayed at that magnitude for the next $\sim$1476 days. A second rise is observed in 2022, beginning 404 days before brightening by a further 2.7 mag above the bright plateau. The second mid-IR outburst could have started 204 days before the optical outburst.


The near-IR $1-2.5~\mu$m spectrum taken during the second outburst shows all of the characteristics of a FUor outburst. The triangular shape in the H-band due to absorption from water vapor bands is clearly present. In addition, we observe strong absorption from $^{12}$CO at $\lambda>2.29\mu$m. The He I line at $1.08 \mu$m, commonly associated with outflows in FUor outbursts, is also seen in absorption.

A previous Gaia-discovered outburst, Gaia17bpi, also displayed a two-stage mid-IR brightening that started $\sim$ 1.5 years before the onset of the optical outburst \citep{2018Hillenbrand}. This could suggest a similar mechanism to that of GPSV16, in which the outburst propagates inward. This type of propagation is suggested in the magnetically activated accretion outburst models of \citet{2023Cleaver}. Here, a YSO would become brighter in the mid-IR years (or decades) before we see a change in the optical. Interestingly, the strongest outburst modeled by \citet{2023Cleaver} led to longer lag times between the mid-IR and the optical. GPSV16 shows a longer lag time ($\sim$8 years) and larger outburst luminosity ($\sim130$ L$_{\odot}$) compared with Gaia17bpi (1.5 years and $\sim8$L$_{\odot}$, respectively).

This work demonstrates the importance of multi-wavelength monitoring of eruptive YSOs over long baselines to understand the physical mechanisms driving their outbursts.

\section*{Acknowledgements}

This research has made use of the NASA/IPAC Infrared Science Archive, which is funded by the National Aeronautics and Space Administration and operated by the California Institute of Technology.
C.C.P. was supported by the National Research Foundation of Korea (NRF) grant funded by the Korean government (MEST) (No. 2019R1A6A1A10073437).
P.W.L. acknowledges support by grant ST/Y000846/1 of the UK Science and Technology Facilities Council.
This work was supported by the New Faculty Startup Fund from Seoul National University and the NRF grant funded by the Korean government (MSIT) (grant numbers RS-2024-00416859 and RS-2026-25490557).  G.J.H. is supported by  general grant 12573031 from the National Natural Science Foundation of
China and by National Key R\&D program 2022YFA1603102
from the Ministry of Science and Technology (MOST) of
China.
D.J.\ is supported by NRC Canada and by an NSERC Discovery Grant.
Z.G. is supported by the China-Chile Joint Research Fund (CCJRF No.2301) and the Chinese Academy of Sciences South America Center for Astronomy (CASSACA) Key Research Project E52H540301. Z.G. was supported by the National Research and Development Agency (ANID) through the FONDECYT Iniciaci\'on project Grant No. 11260176. Z.G. and C.M. are funded by the project ALMA-ANID 31240014
This work has made use of data from the European Space Agency (ESA) mission
{\it Gaia} (\url{https://www.cosmos.esa.int/gaia}), processed by the {\it Gaia}
Data Processing and Analysis Consortium (DPAC,
\url{https://www.cosmos.esa.int/web/gaia/dpac/consortium}). Funding for the DPAC
has been provided by national institutions, in particular the institutions
participating in the {\it Gaia} Multilateral Agreement.
\section*{Data Availability}

The photometric data is available in the article and in its
online supplementary material.
The spectra of GPSV16 will be shared on reasonable
request to the corresponding author.



\bibliographystyle{mnras}
\bibliography{gpsv16} 








\bsp	
\label{lastpage}
\end{document}